\documentstyle[11pt,paspconf,psfig]{article}

\begin{document}

\title{The High-Resolution OTF Survey of the $^{12}CO$ in M\,31}

\author{N. Neininger\altaffilmark{1,2}}
\affil{Radioastronomisches Institut der Universit\"at Bonn (RAIUB), Germany}

\affil{in collaboration with}

\author{Ch. Nieten and R. Wielebinski\altaffilmark{1}}
\altaffiltext{1}{Max-Planck-Institut f\"ur Radioastronomie (MPIfR), 
Bonn, Germany}
\author{M. Gu\'elin and R. Lucas\altaffilmark{2}}
\altaffiltext{2}{Institut de Radioastronomie Millim\'etrique (IRAM), 
Grenoble, France}
\author{H. Ungerechts\altaffilmark{3}}
\altaffiltext{3}{Instituto de Radioastronom\'{\i}a Milim\'etrica (IRAM), 
Granada, Spain}

\begin{abstract}
We are performing a fully sampled survey of the $^{12}$CO(1-0) line
emission of the nearby galaxy M\,31 with the IRAM 30-m telescope.  The
resolution obtained is $23''$ which corresponds to about 90\,pc along
the major axis.  This small beam is well adapted to the size of
typical structures of the molecular gas and allows to detect even
interarm cloud complexes.  We chose an on-the-fly observing technique
which allows us to cover the large area of about 1 square degree in a
reasonable time.

Until now, almost three quarters of the galaxy have been observed in
the (1-0) line and some selected regions with a longer integration
time in the (2-1) transition as well.  Compared to the H{\sc i}
emission, the spiral arms show a higher contrast and more structure,
albeit the general morphology is very similar.  A first analysis
suggests that the CO emission is a very good tracer of the molecular
gas in M\,31.  In addition to the survey, we perform observations of
selected complexes down to the sub-arcsecond level with the IRAM
interferometer.  Together with data of other species such as dust or
neutral carbon we hope to proceed towards a better understanding of
the molecular gas at large {\em and} small scales.
\end{abstract}



\section{Introduction}

The investigation of the large-scale conditions for the star formation
is an important problem of modern astronomy.  The final stage of the
process is reached when individual regions within diffuse molecular
clouds start to coagulate into dense, gravitationally bound complexes
that eventually collapse and form stars.  The trigger to start such a
contraction is however unknown, and in particular the influence of
large-scale phenomena on this process is unclear.  Possible candidates
are cloud-cloud collisions (e.g.\ due to orbit crowding), MHD shocks
(caused by density waves), `contagious' star formation induced by
supernova blast waves or thermal and magnetic instabilities.  The
relative importances of these processes are however uncertain and the
time scales are unknown.

The small-scale properties of star-forming clouds are studied in great
detail in the Milky Way, but there are severe difficulties to obtain
good information about the large-scale structure.  This is of course
due to our position within the disk of the Galaxy, but also because of
the problem to determine proper distances.  We thus have to
investigate an external galaxy and the obvious choice is the Andromeda
Galaxy, M\,31.  It is rather close, at a well determined distance
(784\,kpc after Stanek \& Garnavich 1998), its properties are similar
to that of our Milky Way and a wealth of observations at all
wavelengths is available for comparisons.  In particular, a complete
survey of the $\lambda$ 21\,cm line emission of the atomic hydrogen
has been performed with the WSRT at $24''\times 36''$ angular
resolution (Brinks \& Shane 1984).

Several attempts have been made to observe the emission of the CO in
that galaxy in order to investigate the properties of the molecular
gas, but this turned out to be unexpectedly difficult.  Early attempts
(e.g.\ Combes et al.\ 1977) resulted in a rather high number of
non-detections -- only the dustiest regions showed a reasonable
probability to detect molecular gas (e.g.\ Lada et al.\ 1988, Ryden \&
Stark 1986).  Such a sample is however biased, of course, and only a
survey with uniform coverage can give the necessary information about
the large-scale properties.  The first complete survey of M\,31 was
however published only a few years ago (Dame et al.\ 1993) and had an
angular resolution of $9'$ (2\,kpc along the major axis).

\section{Observations} \label{sec:observations}
\subsection{Preparatory Studies} \label{sec:prep}

In 1993, we made a new attempt to study the properties of an unbiased
sample of molecular clouds and chose the target region on kinematical
grounds only: we focused on a place where the major axis is crossed by
a spiral arm whose location was determined by a kinematical analysis
of the H{\sc i} data (Braun 1991).  The observations were performed
with the IRAM 30-m telescope in a standard ``on-off'' observing mode. 
At a distance of about $38'$ south-west from the center, CO emission
was indeed clearly detected around the major axis and subsequently an
area of about $3'\times 4'$ was mapped.  We found an extended cloud
complex and several individual clumps within this area.  The
properties of the clouds varied substantially within these few
arcminutes, e.g.\ we observed line widths from FWHM 20 km\,s$^{-1}$
down to only 4 km\,s$^{-1}$.  The line temperatures found were of the
order of 0.1 to 0.4 K (in T$_A^*$ units) -- and not limited to only
about 20\,mK as found by Dame et al.\ (1993).

The results showed clearly that the CO emission of M\,31 is highly
clumped and the low intensities of the CfA survey are caused by the
small filling factor of the $9'$ beam.  A high spatial resolution is
thus mandatory to reveal the properties of the molecular gas.  This is
however a difficult task because of the size of the emission area: a
surface of about 1 square degree has to be mapped with an angular
resolution of less than half an arcminute.  The sensitivity should
reach 0.2\,K or better at a velocity resolution of a few km\,s$^{-1}$.

In principle, there are at least two ways to substantially speed up
the imaging speed compared to the classical on-off mode: one could use
multiple-beam receivers and thus cover a larger field with every
``on''-position or use a continuous scanning which reduces overlays
for telescope moving and reference positions by a very large amount.

\subsection{Observation technique} \label{sec:obs}

Based on the experience obtained with the little map and the various
already published results we choose an ``OTF'' (On-The-Fly) observing
technique for our survey.  Here, the telescope is moved across the
source at a constant speed, taking data at a high rate, so that there
is a sufficient number of dumps within the diameter of the beam.  In
our case, we use a scanning speed of $4''$/second and take a dump
every second, which ensures a good sampling of the $22''$ beam.  The
field is covered by adding subsequent scans parallel to the first at a
small distance, in our case $8''$.  The orientation of the scans is
defined in a coordinate system fixed to the center of the source; on
the sky, the scans therefore remain equidistant and the individual
dumps equally spaced within the tracking errors of the telescope ($\ll
1''$ during a normal scan).  That way, the coverage thus obtained is
uniformly and densely sampled.

The reference data are obtained before and after each scan at
positions free of emission; typically, we integrated 30 seconds there,
so that the noise of this data is significantly lower than the noise
of the dumps along the scan.  A scan of a length of $20'$ lasts five
minutes given the parameters above, so a complete sequence
(calibration -- reference -- scan -- reference) takes about 6.5
minutes to complete.  In order to obtain the sensitivity needed, two
SIS receivers are used in parallel which look at the same sky
position.  Moreover, each field is scanned twice, in orthogonal
orientations.  That way, we are able to obtain a sensitivity of $\sim
0.15$ K or better in the final map (T$_A^*$ scale).

The orthogonal orientation of the second coverage not only results in
a dense sampling of the source area, but allows for a special data
reduction technique that supresses so-called ``scanning noise''.  It
is derived from the ``basket-weaving'' method presented by Emerson \&
Gr\"ave (1988) for cm continuum observations and adapted to line
observations by P. Hoernes (1998).  A short overview of the procedure
is given below in Sect.\ \ref{sec:reduc}.  As a result, the noise
distribution is very smooth and in particular we avoid spurious
elongated artifacts.

The main backends used are filterbanks with 1\,MHz resolution.  Three
units are available at the 30-m telescope: two of them offer 256
channels and one 512 channels.  The total bandwidth of the receivers
and the IF system being 500 MHz each, the necessary velocity coverage
for M\,31 is thus easily obtained with a sufficient resolution (2.6
km\,s$^{-1}$).  The 512-channel filterbank and the autocorrelators
also available at the telescope were used for the two 230\,GHz
receivers that run in parallel to the 115\,GHz units.  Each point may
thus be observed with up to four receivers simultaneously -- but for
proper observations at the higher frequency the observational
parameters have to be adapted to the smaller beam, of course.

\subsection{Data reduction} \label{sec:reduc}

The observation method results in separate files for the calibration,
the reference and the source data.  The first step in the data
reduction process is thus to apply the appropriate {\bf calibration}
to each dump.  Thereafter, the spectrum at the {\bf reference}
position is subtracted from the source points.  Various averaging
options for the case of multiple references are available, but a
straight mean of the values obtained before and after the individual
scan is usually sufficient.  After this subtraction, we have a set of
spectra comparable to the usual output of ``on-off'' or wobbler
techniques, with one second of integration time per spectrum.

The next step in the data reduction is the subtraction of spectral
{\bf baselines} and the removal of erroneous data points.  Such points
are e.g.\ introduced by bad channels in the filterbanks.  They are
relatively easy to detect due to their ``singular'' nature and the
occurrence at the same channel in independent spectra.  The fitting of
the baseline needs a bit more thought: since each scan contains
several hundred spectra, we need a routine that is able to determine
the necessary line windows by itself.  Here, we profit largely from
the WSRT H{\sc i} data that have a resolution comparable to our survey
($24''\times36''$, Brinks \& Shane 1984).  We assume that the velocity
structure of the atomic gas (H{\sc i}) is similar to that of the
molecular gas (CO).  Then, we can use the kinematical information in
the H{\sc i}-data to determine the velocity interval possibly
containing CO emission and fit the spectral baseline to the data
outside of this region.  In the case of M\,31 this is a very safe
procedure because the CO emission is weak and even a possible error in
the line window does not change the fit of the baseline very much.  In
any case, the baseline fit is a linear procedure and may be used
iteratively, if necessary.

Now it is time to actually `reduce' the data -- remember that we still
are working on a $4''\times 8''$ data grid for the $22''$ beam.  We
thus {\ bf regrid} the data onto a convenient regular grid, which at
the same time allows to correct for tracking errors and to introduce
additional scans (e.g.\ to back up single lines with higher noise). 
It may be advisable, however, to maintain the distance between
individual scans in the grid setup for the next step of the data
reduction.

\begin{figure}
    \vbox{\vbox{\vbox{
    \psfig{file=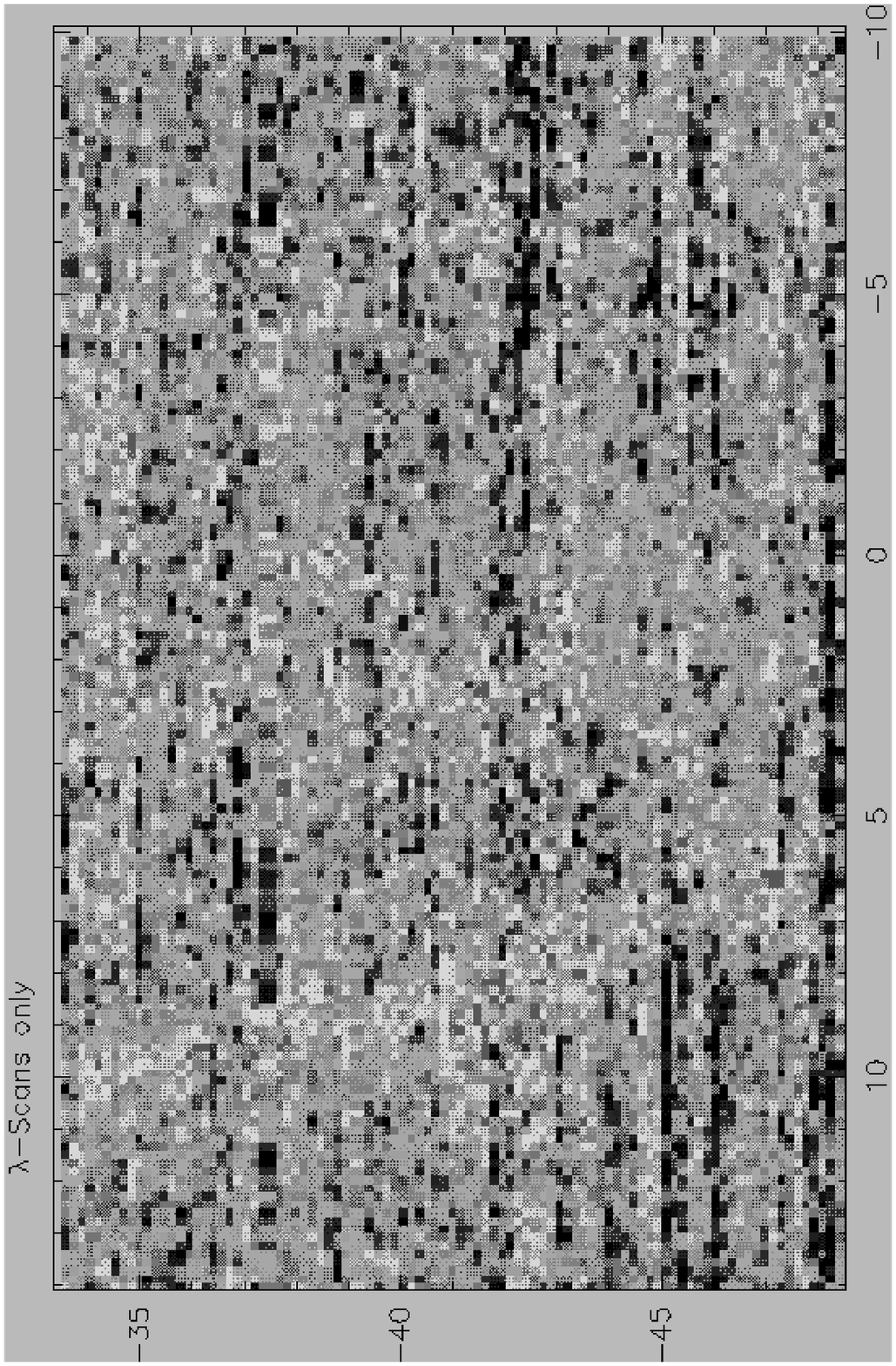,width=60mm,angle=270}
    \vspace{-38.7mm}} \hfill
    \psfig{file=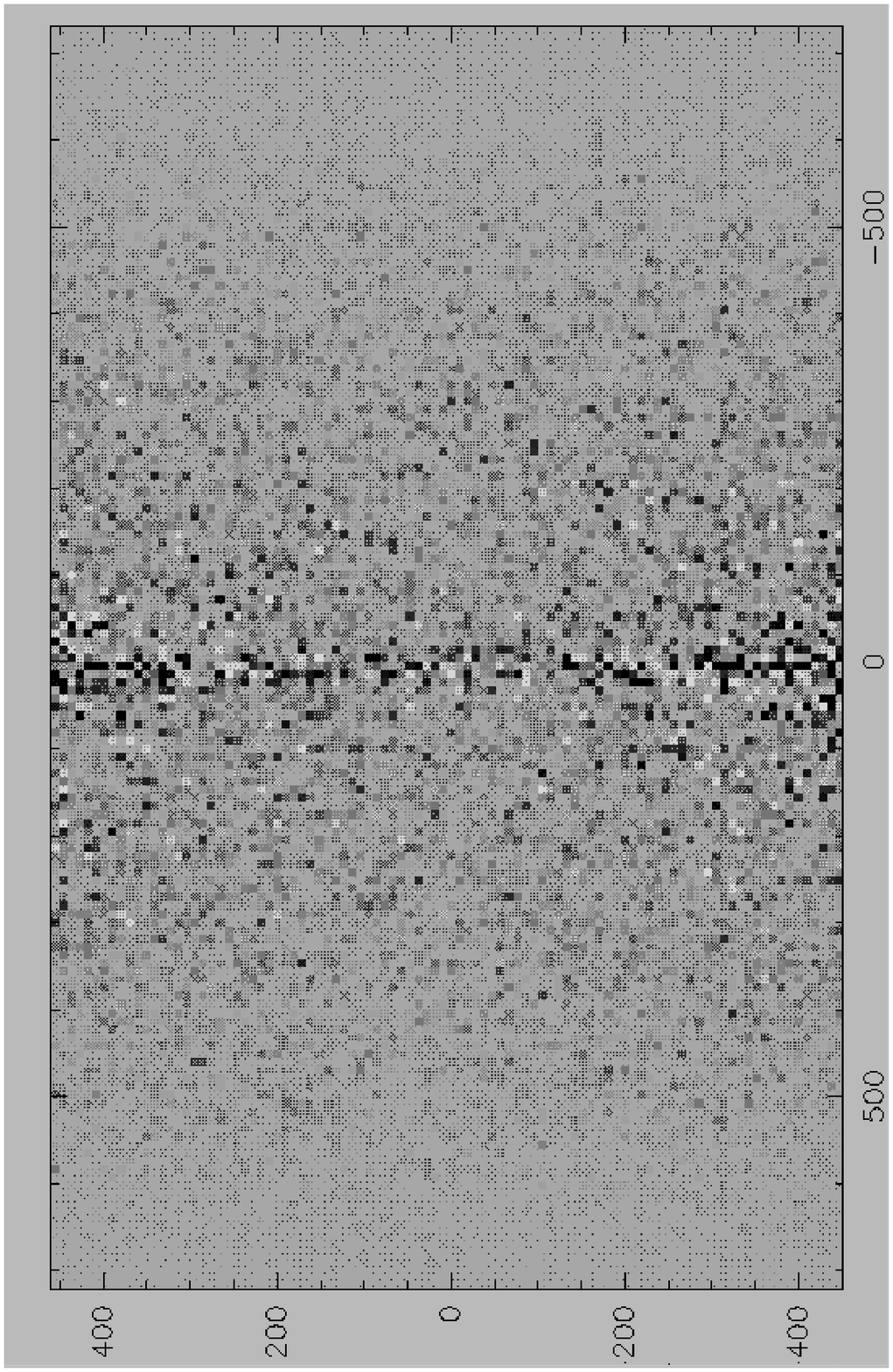,width=60mm,angle=270}
    \vspace{6mm}}
    \psfig{file=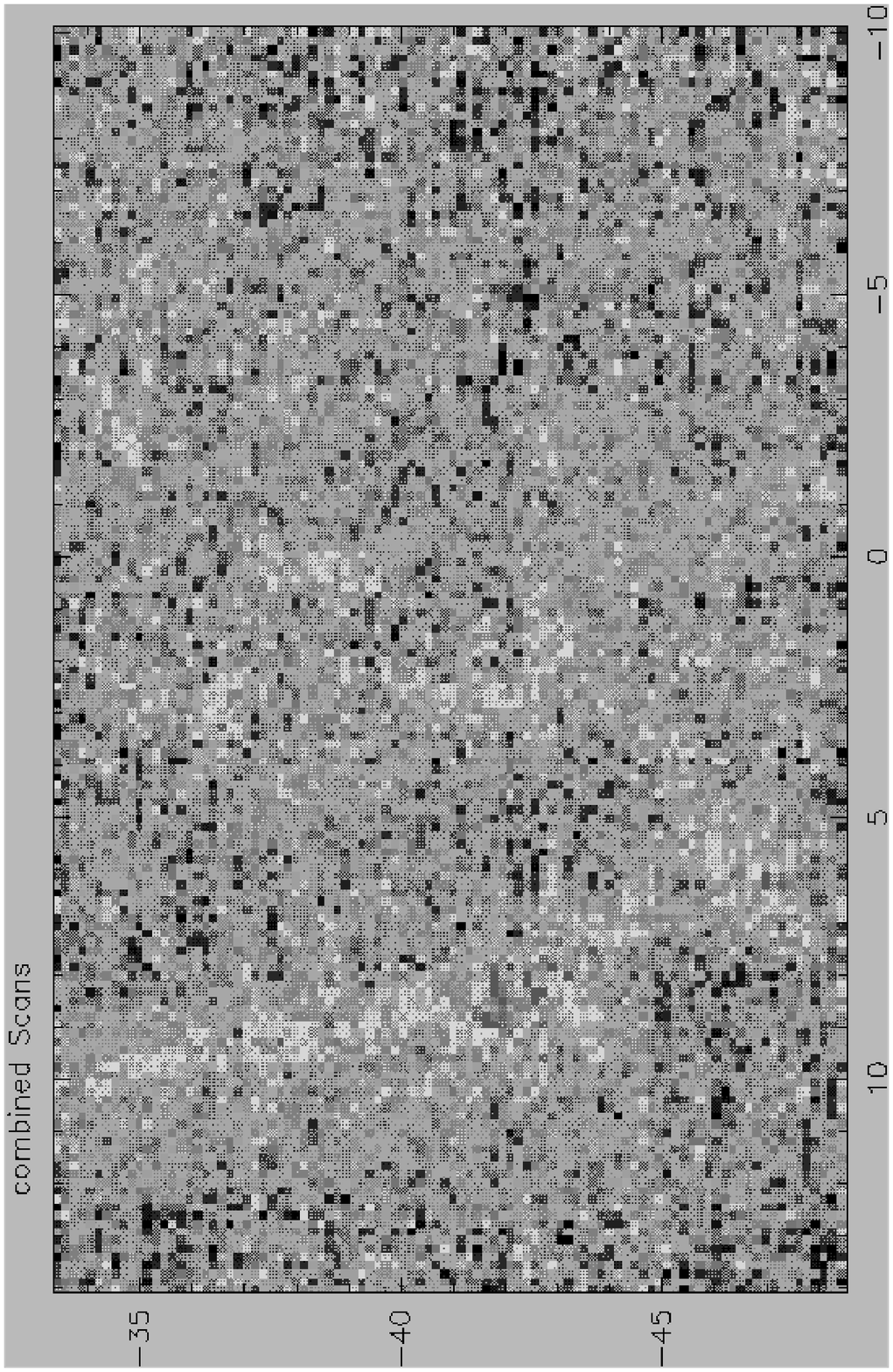,width=60mm,angle=270}
    \vspace{-44.0mm}}\hfill
    \parbox{75mm}{\caption{Example of the FFT filtering: in the 
    upper row a raw image is shown, scanned in one direction only 
    (left), together with its Fourier transform (right). The stripes 
    are transformed into a very narrow strip which is weighted down.
    To the left, the result of the filtering and combination 
    is given. Only random noise remains.}} \label{fig:fft}
\end{figure}

At this stage, we {\bf combine} several maps (which are actually data
cubes) using the ``basket-weaving'' method already mentioned above
(Emerson \& Gr\"ave 1988).  Inspection of the maps shows that quite
often drifts of the zero level are remaining even after the
subtraction of a spectral baseline.  Adjacent scans thus may show a
different value for neighboring points (at a distance of $8''$
compared to a beam of $22''$) because they have been observed through
a different atmosphere, separated in time by e.g.\ six minutes (see
Fig.\ \ref{fig:fft} upper left).  Now remember that we have scanned
the area of interest at least twice, at orthogonal orientations. 
Thus, we have two data sets where the distribution of coherent dumps
(along the scan) and incoherent dumps (in different scans) is
orthogonal.

In the two-dimensional frame of a map, the scans constitute regularly
spaced rows or columns.  Any feature that occurs with such a fixed
period (e.g.\ a varying zero level in each scan) will be projected
into a very narrow interval by a 2-D Fourier transform (Fig.\
\ref{fig:fft}, upper right).  Thus, it is easy to apply a filter
function to suppress this noise contribution.  The orientation of this
interval depends of course on the original scanning direction, so the
part taken out can be reconstructed from the other input maps.  This
also avoids unwanted filtering of linear source structures, because
{\em a priori} only periodic arrangements are filtered -- and only in
one specific orientation per input frame.

After the filtering, the input maps are coadded -- in principle, any
number of coverages with arbitrary scanning orientations can be used
as input.  If the coverage should be not the same for all input maps,
the quality of the noise filtering will vary over the combined area,
of course.  However, the procedure is rather `friendly' and does not
introduce edge-effects at the borders of the individual coverages. 
For spectral data, each spectral channel is treated as an individual
map.  The whole procedure works on the data pixels and it is up to the
user to ensure a coherent data set beforehand.

Until now, the pixels are coherent only at the level of the gridding,
which is typically of the order of half a beam width or less.  Hence,
we {\bf smooth} the data as the last step of the data reduction.  The
smoothing gaussian is chosen small enough in order not to degrade the
original resolution (we obtain a final resolution of $23''$); this
leaves the source structures untouched, but further suppresses the
random noise in the pixels.

\section{Results} \label{sec:results}

\subsection{The map} \label{sec:map}

\begin{figure}
    \psfig{file=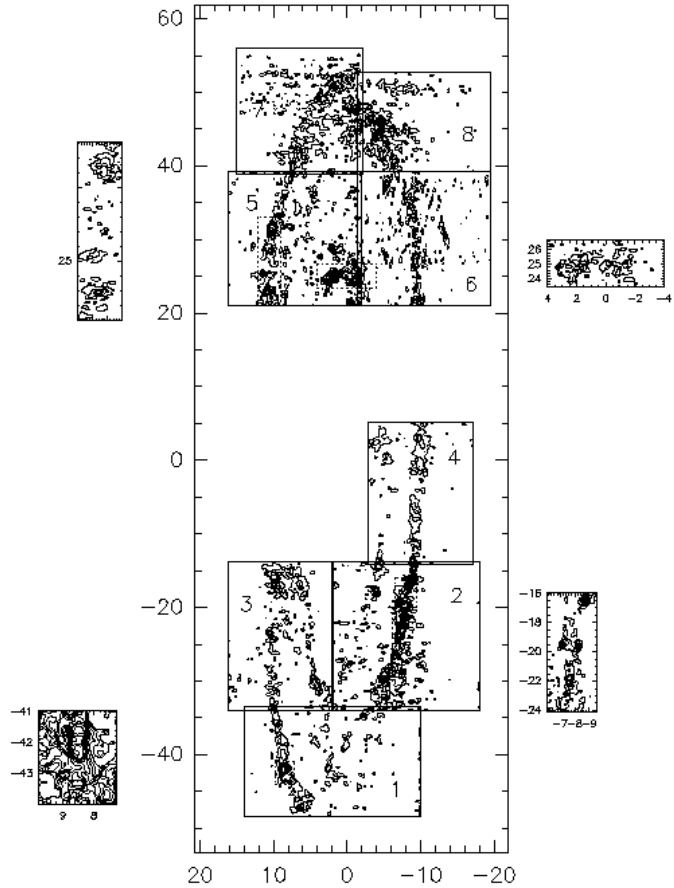,width=120mm,angle=0} 
    \caption{Overview of the mapped regions as of summer `99.  The
    coordinates are fixed at the center of M\,31 (R.A.$_{B1950.0}$
    $0^h40^m00.3^s$, Dec.$_{B1950.0}$ $41^{\circ}00'03''$); X is positive
    towards the NE along the major axis (P.A. $37.7^{\circ}$), Y is
    positive to the SE along the minor axis.  The smaller fields show
    selected regions that have been mapped with a full sampling in the
    (2-1) line as well.  Note that part of the data in the northern part
    is still preliminary and the contours have been chosen such as to show
    some effects of the scanning noise.} \label{fig:gesamt}
\end{figure}

Until now, about three quarters of M\,31 have been mapped with full 
sampling for the $22''$ beam at the $^{12}$CO(1-0) line transition 
(see Fig.\ \ref{fig:gesamt}). In addition, we choose several regions 
with particular properties and observed them in a mode adapted to 
fully sample the smaller beam at the (2-1) transition (shown in the 
fields surrounding the main frame in Fig.\ \ref{fig:gesamt}).

Over most of the area surveyed so far, coherent spiral arm pieces are
clearly visible.  The arm/interarm contrast is high, but there are
small cloud complexes scattered over a large part of the surface. 
Note however, that only in the southern half of the map the lowest
contour is at least $3\sigma$, whereas in the northern part the data
was still preliminary at the time of the conference.  There, the
contours have been chosen such as to show the effects of the scanning
procedure which result e.g.\ in the horizontal stripes at the upper
end of fields 7 and 8.  In any case, we check such weak points with
dedicated on-off measurements and maintain only those that can be
confirmed -- which is usually the case.

\subsection{Comparisons with other data and first results} \label{sec:resu}

There is a wealth of observations available for M\,31, covering the
whole wavelength range.  As described above, we used the H{\sc i} data
cube of the WSRT survey (Brinks \& Shane 1984) already for the
determination of the line windows during the data reduction.  Now, we
can compare the two data sets to obtain a picture of the properties of
the atomic vs.\ the molecular gas.  The molecular gas turns out to
show a significantly higher contrast, but the general structure is
very similar to that of the atomic gas: the main emission is found in
a ``ring'' and very few gas can be found close to the center.  The
ratio of molecular/atomic gas decreases with radius, but individual
molecular complexes have nevertheless been found out to a distance of
about 18\,kpc.  The kinematical signature is rather similar for both
kinds of gas, which justifies the use of the H{\sc i}data in the
baseline fitting procedure.

\begin{figure}
    \psfig{file=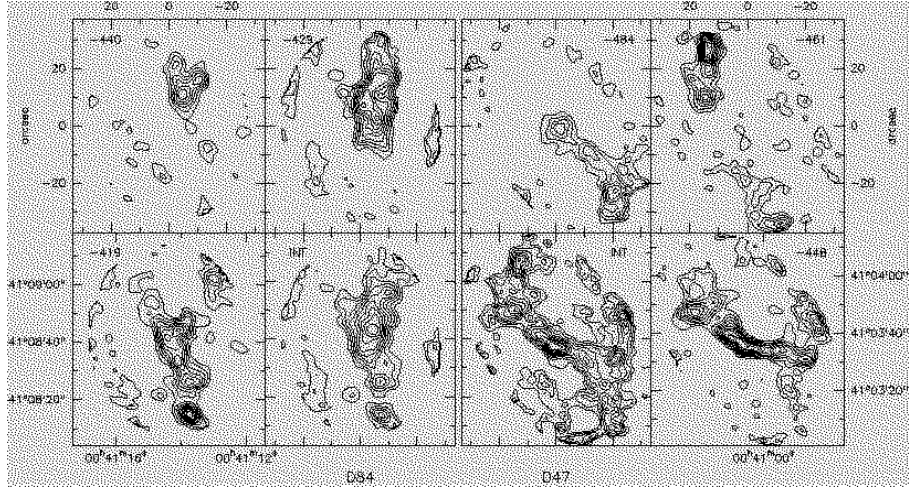,width=120mm,angle=270}
    \caption{An example of two cloud complexes with different 
    properties, observed with the PdB interferometer. The left half 
    shows the quiescent complex D84, the right half the disturbed 
    region D47. The two lower inner panels (labeled `INT') give the 
    integrated emission, whereas the outer panels show the structure 
    in selected velocity bins, labeled with the central velocity. Note 
    the absence of emission at the central position of D47 in the 
    $-461$ km/s channel; the corresponding spectrum is double-peaked 
    with a 40\,km/s wide gap.} \label{fig:pdb}
\end{figure}

Large-scale streaming motions are not prominent along the spiral 
arms -- the data are rather dominated by local effects. They show up 
as double- or multiple-peaked spectra with total widths of up to 50 
km\,s$^{-1}$ (cf.\ Fig.\ \ref{fig:pdb}). A comparison with the 
distribution of ionized regions (Devereux 1994) suggests a relation 
between such disturbed molecular clouds and the H{\sc ii} regions (see 
Fig.\ 2 of Neininger et al.\ 1998) for the region presented in Fig.\ 
\ref{fig:pdb}).

\section{Summary}

\paragraph{Technical items} The OTF method is a fast, flexible and
versatile observing mode which yields high-quality data.  Longer
integration times per point can be achieved by co-adding a
corresponding number of coverages.  In principle, it can be adapted to
all mapping projects with single-dish telescopes -- the limitations
are usually technical, such as receiver stability, maximum telescope
speed, maximum dump rate or data storage capacity.  Our setup allows
us to map an area of 10\,arcmin$^2$ at a typical noise of 150\,mK per
spectrum in 1 hour of telescope time.  Using specially designed
filtering techniques, spurious features can be removed and a
homogeneous result can be achieved -- even in wavelength bands that
are seriously affected by atmospheric effects.

\paragraph{Astronomical results} The molecular gas in M\,31 has a high
arm/interarm contrast, but single cloud complexes are found between
the arms.  The spatial filling factor is rather low.  The bulk of the
molecular gas is situated between a radius of about 4\,kpc and
13\,kpc, but some cloud complexes have been observed out to distances
of 18\,kpc.  The total gas mass (H{\sc i} + H$_{2}$) corresponds well
to the optical extinction.  The signature of a possible density wave
is very weak in the molecular gas, instead it seems to be dominated by
local effects.

\acknowledgments Special thanks to the IRAM staff which has made the
OTF observations possible -- in particular to A. Sievers, W. Brunswig,
W. Wild and the receiver engineers.  Of special importance were some
hydrogen data: we thank E. Brinks and R. Braun for the H{\sc i} cubes
and N. Devereux for his H$\alpha$ image.

%
%

%

\end{document}